\documentclass[prd,aps,floats,floatfix,eqsecnum,nofootinbib]{revtex4}
\usepackage{amsmath,amssymb,verbatim,epsfig,psfig,graphicx,rotating}
\usepackage{psfrag}
\newcommand{\be}{\begin{equation}}
\newcommand{\ee}{\end{equation}}
\newcommand{\bea}{\begin{eqnarray}}
\newcommand{\eea}{\end{eqnarray}}
\begin{document}
\title{Clarifying inflation models: slow-roll as an expansion in
$1/N_{efolds}$.}
\author{\bf D. Boyanovsky$^{(c,a)}$} \email{boyan@pitt.edu}
\author{\bf H. J. de Vega$^{(b,a)}$}\email{devega@lpthe.jussieu.fr}
\author{\bf N. G. Sanchez $^{(a)}$}\email{Norma.Sanchez@obspm.fr}
\affiliation{$^{(a)}$
Observatoire de Paris, LERMA, Laboratoire Associ\'e au CNRS UMR 8112,
 \\61, Avenue de l'Observatoire, 75014 Paris, France. 
\\$^{(b)}$ LPTHE, Laboratoire Associ\'e au CNRS UMR 7589,\\
Universit\'e Pierre et Marie Curie (Paris VI) et Denis Diderot (Paris VII),\\
Tour 24, 5 \`eme. \'etage, 4, Place Jussieu, 75252 Paris, Cedex 05,
France.}
\affiliation{$^{(c)}$Department of Physics and
Astronomy, University of Pittsburgh, Pittsburgh, Pennsylvania 15260,
USA}

\begin{abstract}
Slow-roll inflation is studied as an effective field theory. We
find that the form of the inflaton potential consistent with WMAP
data and slow roll is $ V(\phi) = N \; M^4 \;
w\left(\frac{\phi}{\sqrt{N} \; M_{Pl}}\right) $ where $ \phi $ is
the inflaton field, $ M $ is the inflation energy scale,   and $ N
\sim 50 $ is the number of efolds since the cosmologically
relevant modes crossed the Hubble radius until the end of
inflation. The inflaton field scales as $ \phi = \sqrt{N} \;
M_{Pl} \; \chi $. The dimensionless function $ w(\chi) $ and field
$ \chi $ are generically $ \mathcal{O}(1) $. The WMAP value for
the amplitude of scalar adiabatic fluctuations $
|{\Delta}_{k\;ad}^{(S)}|^2 $ fixes the inflation scale $M \sim
0.77\times 10^{16}$. This   form of the potential   makes manifest
that the slow-roll expansion is an expansion in $1/N$.  A
Ginzburg-Landau realization of the slow-roll inflaton potential
reveals that the Hubble parameter, inflaton mass and non-linear
couplings are of the see-saw form in terms of the small ratio $
M/M_{Pl} $. For example, the quartic coupling $ \lambda \sim
\frac{1}{N} \left(\frac{M}{M_{Pl}}\right)^4 $. The smallness of
the non-linear couplings is {\bf not} a result of fine tuning but
a {\bf natural} consequence of the validity of the effective field
theory and slow roll approximation. We clarify Lyth's bound
relating the tensor/scalar ratio and the value of $ \phi/M_{Pl} $.
The effective field theory is valid for $ V(\phi)
\ll M_{Pl}^4 $ for general inflaton potentials allowing amplitudes
of the inflaton field $ \phi $ {\bf well beyond} $ M_{Pl} $. Hence
bounds on $r$ based on the value of $\phi/M_{Pl}$ are overly
restrictive. Our observations lead us to suggest that slow-roll,
single field inflation may well be described by an almost critical
theory, near an infrared stable gaussian fixed point.
\end{abstract}

\date{\today}
\maketitle
\tableofcontents

\section{Introduction and Results}

Inflation was originally proposed to solve several outstanding
problems of the standard Big Bang model
 \cite{guth,kolb,coles,lily,riottorev} thus becoming an important
paradigm in cosmology. At the same time, it provides a natural
mechanism for the generation of scalar density fluctuations that
seed large scale structure, thus explaining the origin of the
temperature anisotropies in the cosmic microwave background (CMB),
as well as that of  tensor perturbations (primordial gravitational
waves).
 A distinct aspect of
inflationary perturbations is that these are generated by quantum
fluctuations of the scalar field(s) that drive inflation. After
their wavelength becomes larger than the Hubble radius, these
fluctuations are amplified and grow, becoming classical and
decoupling from  causal microphysical processes. Upon re-entering
the horizon, during the matter era, these classical perturbations
seed the inhomogeneities which generate structure upon gravitational
collapse. While there is a great
diversity of   inflationary models, most of them predict fairly
generic features: a gaussian, nearly scale invariant spectrum of
(mostly) adiabatic scalar and tensor primordial fluctuations. These
generic predictions of most inflationary models make the
inflationary paradigm fairly robust.

Inflationary dynamics is typically studied by treating  the
inflaton as a homogeneous  classical scalar
field \cite{kolb,coles,lily,riottorev} whose evolution is
determined by a classical equation of motion, while the quantum
fluctuations of the inflaton  provide the seeds for the scalar
density perturbations of the metric and are treated in the
Gaussian approximation. In quantum field theory, the classical inflaton
corresponds to the expectation value of a quantum field operator in a
translational invariant state.

Although there is a wide variety of inflationary models, the
WMAP \cite{WMAP} data can be fit outstandingly well by simple
single field slow roll models.

Inflation based on a scalar inflaton field should be considered as
an {\bf effective theory}, that is,  not necessarily a fundamental
theory but as a low energy limit of a microscopic fundamental
theory. The inflaton may be a coarse-grained average of
fundamental scalar fields, or a composite (bound state) of fields
with higher spin, just as in superconductivity. Bosonic fields do
not need to be fundamental fields, for example they may emerge as
condensates of fermion-antifermion pairs $ < {\bar \Psi} \Psi> $
in a grand unified theory (GUT) in the cosmological background. In
order to describe the cosmological evolution is enough to consider
the effective dynamics of such condensates.  The relation between
the low energy effective field theory of inflation and the
microscopic fundamental theory is akin to the relation between the
effective Ginzburg-Landau theory of superconductivity and the
microscopic BCS theory, or like the relation of the $O(4)$ sigma
model, an effective low energy theory of pions, photons and chiral
condensates with quantum chromodynamics (QCD) \cite{quir}. The
guiding principle to construct the effective theory is to include
the appropriate symmetries \cite{quir}. Contrary to the sigma model
where the chiral symmetry strongly constraints the
model \cite{quir}, only general covariance can be imposed to the
inflaton model.

While inflationary cosmology is currently studied from the point
of view of classical field theory with small quantum corrections,
non-perturbative quantum aspects of the dynamics of inflation were
studied in refs.\cite{cosmo,cosmo2,staro,woodard}. More recently
particle decay in a de Sitter background as well as  during slow
roll inflation has been studied in ref.\cite{pardec} together with
its implication for the decay of the density fluctuations. Quantum
corrections to slow roll inflation including quantum corrections
to the effective inflaton potential and its equation of motion are
derived in ref.\cite{qua}.

Recent studies of quantum corrections during
inflation \cite{pardec,qua} revealed the robustness of classical
single field slow roll inflationary models as  result of the
validity of the effective field theory description. The reliability
of an effective field theory of inflation hinges on a wide
separation between the energy scale of inflation, determined by $H$
and that of the underlying microscopic theory which is taken to be
the Planck scale $M_{Pl}$.

The data from WMAP provides an upper bound on the scale of the
inflationary potential \cite{WMAP} $V^{1/4} < 3.3 \times
10^{16}~\textrm{GeV}$ ($95 \% CL$), thereby establishing un upper
bound on the scale of inflation $ H < 2.6 \times 10^{14}\,\textrm{GeV} $.
Hence, the smallness of the ratio $ H/M_{Pl} \lesssim 10^{-4} $ warrants
the reliability of the effective field theory approach. A simple
Ginzburg-Landau realization of the inflationary potential as an
effective field theory has been recently shown to fit the WMAP
data remarkably well \cite{nos}.

As mentioned in refs.\cite{pardec,qua} there are \emph{two
independent} expansions: the effective field theory (EFT) one
based on the small dimensionless ratio $H/M_{Pl}$ and the slow
roll expansion. The latter one is an \emph{adiabatic} expansion
which relies on a fairly flat inflationary potential and invokes a
hierarchy of dimensionless ratios that involve derivatives of the
inflationary potential \cite{barrow,lily,WMAP,hu}.

\medskip

{\bf The goal of this article:}

In this article we combine the results from WMAP and the slow roll
expansion to suggest that the inflationary potential has a
\emph{universal} form which helps to clarify  both the (EFT) and
slow roll expansions. The main point of the argument is  the
presence of two independent small parameters in single field
inflationary cosmology: $ H/M_{Pl} $ and $ 1/N $, where $N$ is the
number of e-folds before the end of inflation during which
wavelengths of cosmological relevance today first cross the
horizon. Consistent inflationary models require that $N \sim
50-60$. We show that the form of the potential suggested by the
data and slow roll leads naturally to the identification of the
slow-roll expansion as an expansion in $1/N$.

While the (EFT) ratio $H/M_{Pl}$ and the slow roll parameters are
logically independent, slow roll implies a large number of
e-folds, therefore the hierarchy of slow roll parameters is
related to the smallness of $1/N$. While this point is widely
known and understood, our main observation is that the slow roll
expansion is a \emph{systematic} expansion in the small parameter $1/N$.

\bigskip

{\bf Brief summary of results:}
\begin{itemize}
\item{ We observe that combining the WMAP data with the
 slow roll expansion  \emph{suggests} a \emph{consistent} description
 of single field  inflation in terms of a classical potential of the form
\be \label{V}
V(\phi) = N \; M^4 \; w(\chi)
\ee
\noindent where
$w(\chi) \sim \mathcal{O}(1)~,~N\sim 50$  and  $\chi$ is a
dimensionless, slowly varying field \be\label{chifla} \chi =
\frac{\phi}{\sqrt{N} \;  M_{Pl}} \ee \noindent The WMAP data
constrains $M$ to be at the  grand unification (GUT) scale
$\label{M} M \sim 0.77 \times 10^{16}\textrm{GeV} $, which
suggests a  connection between inflation and the physics at the
GUT scale in a cosmological space-time.

\item{The dynamics of the rescaled field $\chi$ exhibits the slow
time evolution in terms of the \emph{stretched} (slow)
dimensionless time variable, \be \label{tau} \tau =  \frac{t \;
M^2}{M_{Pl} \; \sqrt{N}} \ee } The form of the potential and the
rescaled dimensionless field and time variable lead consistently
to slow-roll as an expansion in powers of $1/N$.  }

\item{The inflaton mass around the minimum is given by a see-saw formula
$$
m = \frac{M^2}{M_{Pl}}  \sim 2.45 \times 10^{13} \, \textrm{GeV} \; .
$$
The Hubble parameter when the cosmologically relevant modes exit the horizon
is given by
$$
H  = \sqrt{N} \; m \, h \sim 1.0 \times 10^{14}\,\textrm{GeV}
= 4.1 \; m \; ,
$$
using  $ h \sim 1 $. As a result, $m\ll M$ and $ H \ll M_{Pl}$. A
Ginzburg-Landau  realization of the inflationary potential that
fits the WMAP data remarkably well \cite{nos}, reveals that the
Hubble parameter, the inflaton mass and non-linear couplings are
see-saw-like, namely  powers of the ratio $M/M_{Pl}$ multiplied by
further powers of $1/N$. Therefore, their smallness is not a result
of fine tuning but   a  {\bf natural} consequence of the form of
the potential and the validity of the effective field theory
description and slow roll. The quantum expansion in loops is
therefore a double expansion on $ \left(H/M_{Pl}\right)^2 $ and $
1/N $. Notice that graviton corrections are  also at least of
order $ \left(H/M_{Pl}\right)^2 $ because the amplitude of tensor
modes is or order $H/M_{Pl}$ . While the hierarchy between the
amplitude of the field, $H$, and $M_{Pl}$ is known, we show that
the form of the potential that fits the WMAP data and is
consistent with slow roll suggests the above result for the
non-linear couplings. }

\item{We discuss critically Lyth's bound on the ratio $r$ of the
amplitudes of tensor to curvature perturbations. This bound has
been often used to assess
the feasibility of detection of tensor modes in forthcoming
searches \cite{lyth,efs,vpj}. Since $ r $ can be related to the
change of the inflaton field while the cosmologically relevant
modes exit the horizon \cite{lyth}, the restriction to models with
$ \frac{\phi}{M_{Pl}} \lesssim 1 $  \cite{lyth,efs} yield very
small values of  $ r $. We show that Lyth's bound is overly
restrictive by making precise  the regime of validity of the
effective field theory approach: the use of the inflaton potential
$ V(\phi) $ is consistent for $ V(\phi) \ll M_{Pl}^4 $. This
inequality is fulfilled for values of the inflaton field $\phi$
{\bf well beyond} the Planck mass allowing $ r \gtrsim 1$. We
provide elementary, yet illuminating examples of potentials that
help precise these statements.  Furthermore, the true value of $ r
$ may be very well much smaller than the present upper bound $ r
\lesssim 0.16 $ from WMAP  \cite{WMAP}.

If for whatever reason, a restriction to values for $ \chi
\lesssim 1 $ is invoked, then an {\bf improved} Lyth bound emerges
from our analysis. This improved bound allows  values for $ r $
larger than the previous bounds due to the presence of the new
factor $ \sqrt{N} \sim 7 $  \cite{lyth,efs}.}

 \item{For polynomial realizations of the inflationary potential,
our analysis yields that the non-linear
 couplings scale with inverse powers of $1/N$. In an expanding
 cosmology the logarithm of the scale factor is the number of
 e-folds, thus these couplings scale as powers of $1/\ln(a)$. In a
 quantum field theory, the behavior of the couplings under a change of
scale is dictated by the renormalization group. For example,
 at one-loop level, the quartic coupling of
the $ \Phi^4 $ model in four dimensional euclidean (flat) space
precisely scales as $ 1/\log a = 1/N $,  approaching the trivial
(zero coupling) infrared stable (critical) point when $ a \to
\infty $. Hence, it is \emph{suggestive} that the underlying
reason for the scaling of the couplings in powers of $1/N$ in the
potential eq.(\ref{V}) (see eqn. \ref{Gn} below) may be the
infrared renormalization group running (RG) of long wavelength
modes. This observation leads us to \emph{conjecture} that slow
roll inflation strongly suggests that the effective field theory
is near (but not exactly) a trivial Gaussian infrared fixed point during 
the stage in which scales of cosmological relevance today
crossed the Hubble radius. In this interpretation the theory hovers near
the gaussian fixed point with an almost scale invariant spectrum of scalar 
fluctuations during the slow roll stage, but eventually it must move away from the
neighborhood of this fixed point by the end of inflation to reach the
standard radiation dominated stage.}

\end{itemize}

Eq.(\ref{V}) for the inflaton potential resembles (besides the
factor $ N $) the moduli potential arising from supersymmetry
breaking \cite{riottorev},
\be\label{susy}
V_{susy}(\phi) = m_{susy}^4 \; v\!\left(\frac{\phi}{M_{Pl}}\right) \; ,
\ee
where $ m_{susy} $ stands for the supersymmetry breaking scale. In our
context, eq.(\ref{susy}) indicates  that $  m_{susy} \simeq M
\simeq M_{GUT} $. That is, the susy breaking scale $ m_{susy} $
turns out to be at the GUT and inflation scales. This may be a
\emph{first} observational indication of the presence of
supersymmetry.

\bigskip

In summary, we find that a form of the inflaton potential for
single field models consistent with the WMAP data and slow-roll
is given by eq.(\ref{V}). There is no requirement of fine tuning small
parameters, as these appear in in $ w(\chi) $ naturally in terms of the 
effective field theory ratio  $ \left(H/M_{Pl}\right)^2 $
and $ 1/N $. The form of the potential
eq.(\ref{V}) {\bf encodes } the energy scale of inflation as well
as  slow-roll, and is valid for practically {\bf all} slow-roll
inflaton potentials. In particular, eq.(\ref{V}) applies for all
polynomial potentials investigated in \cite{nos}.

\section{Basic Mass Scales in the Inflaton Potential and the Number of e-folds}

The description of cosmological inflation is based on an isotropic
and homogeneous geometry, which assuming flat spatial sections is
determined by the invariant distance
\be
ds^2= dt^2-a^2(t) \; d\vec{x}^2 \label{FRW} \; .
\ee
The scale factor obeys the Friedman equation
\be \label{ef}
\left[ \frac{1}{a(t)} \; \frac{da}{d t} \right]^2 =
\frac{\rho( t)}{3 \; M^2_{Pl}}   \;  ,
\ee
where $M_{Pl}= 1/\sqrt{8\pi G} = 2.4\times 10^{18}\,\textrm{GeV}$.
In single field inflation the energy density is dominated by a
homogeneous scalar \emph{condensate}, the inflaton, whose dynamics
can be described by an  \emph{effective} Lagrangian
\be\label{lagra}
{\cal L} = a^3({ t}) \left[ \frac{{\dot
\phi}^2}{2} - \frac{({\nabla \phi})^2}{2 \;  a^2({ t})} - V(\phi) \right]
\; .
\ee
\noindent The inflaton potential $ V(\phi) $ is a slowly varying
function of $ \phi $ in order to permit a slow-roll solution for
the inflaton field $ \phi( t) $.

Slow roll inflation corresponds to a fairly flat potential and the
slow roll approximation invokes a hierarchy of dimensionless
ratios in terms of the derivatives of the potential.
Some \cite{barrow,lily,WMAP} of these (potential) slow roll
parameters are given by\footnote{We follow the definitions of
$\xi_V;\sigma_V$ in ref.\cite{WMAP}. ($\xi_V;\sigma_V$ are called
$\xi^2_V;\sigma^3_V$, respectively, in \cite{barrow}).}
\bea
&&\epsilon_V  = \frac{M^2_{Pl}}{2} \;
\left[\frac{V'(\phi)}{V(\phi)} \right]^2  \quad , \quad \eta_V
= M^2_{Pl}  \; \frac{V''(\phi)}{V(\phi)}\; ,
\cr \cr
&& \xi_V = M^4_{Pl} \; \frac{V'(\phi) \; V'''(\phi)}{V^2(\phi)}
\quad , \quad \sigma_V = M^6_{Pl} \;
\frac{[V'(\phi)]^2 \; V^{(IV)}(\phi)}{V^3(\phi)} \; . \label{sig}
\eea
\noindent where $V(\phi)$ is the inflaton potential and primes
stand for derivative with respect to the inflaton field.  The slow
roll approximation \cite{barrow,lily,WMAP,hu} corresponds to
$\epsilon_V \sim \eta_V \ll 1$  with the hierarchy $\xi_V \sim
\mathcal{O}(\epsilon^2_V)~;~\sigma_V \sim
\mathcal{O}(\epsilon^3_V)$, namely $\epsilon_V$ and $\eta_V$ are
first order in slow roll, $\xi_V$ second order in slow roll, etc.

As stressed in ref.\cite{nos}, in order to reproduce the CMB data,
the inflationary potentials in the slow roll scenarios  must have
the structure \be \label{vchi} V(\phi) =  M^4 \;
v\!\left(\frac{\phi}{M_{Pl}}\right) \; , \ee where $v$ is
dimensionless and $M$ determines the energy scale of the potential
during inflation. The minimum of the potential $\phi_{min}$ is
chosen such that  $ V(\phi_{min}) = 0 $ in order to ensure that
inflation ends after a finite number of e-folds. This form of the
potential is also suggested by the dimensionless slow roll
parameters, because the combination $\phi/M_{Pl}$ in the argument
of the potential cancels the ubiquitous factors $M_{Pl}$ in the
slow roll variables.

During slow roll inflation the number of e-folds  before the end
of inflation  at which   the value of the scalar field is
$\phi_{end}$, is given by
\be\label{Nfo} N[\phi(t)] =
-\frac1{M_{Pl}^2} \; \int_{\phi(\tau)}^{\phi_{end}} V(\phi) \;
\frac{d\phi}{dV} \; d\phi \; .
\ee
It is convenient to introduce
$N \sim 50$ as the typical scale of e-folds during which
wavelengths of cosmological relevance today first cross the
horizon during inflation, namely \be \label{N50} 50 =
-\frac1{M_{Pl}^2} \int_{\phi_{50}}^{\phi_{end}} V(\phi) \;
\frac{d\phi}{dV} \; d\phi \; . \ee where $\phi_{50}$ is the value
of the scalar field $50$ e-folds before the end of inflation.

The form of the potential eq.(\ref{vchi}) combined with the above
definition, strongly suggests the following re-scaling of the
field and the potential
\be \label{wchi}
\phi  =  \sqrt{N} \; M_{Pl} \; \chi \qquad , \qquad
V(\phi)  =  N \; M^4 \;  w(\chi) \; ,
\ee
where $\chi$ is dimensionless.  With this definition, the expression
(\ref{N50}) simply becomes
\be \label{uno}
1 = -\int_{\chi_{50}}^{\chi_{end}} \; \frac{w(\chi)}{w'(\chi)} \; d\chi \; ,
\ee
where the prime stands for the derivative with respect to the argument.
The dimensionless field $ \chi $ is \emph{slowly} varying during the stage
of slow roll inflation: a large amplitude change in the field $\phi$
results in a small amplitude change in $\chi$,
\be \label{slofield}
\Delta \chi = \frac{1}{\sqrt{N}} \frac{\Delta \phi}{M_{Pl}}  \; ,
\ee
a change in $\phi$ with $\Delta \phi \sim
M_{Pl}$ results in a change $\Delta \chi \sim 1/\sqrt{N}$. The
number of e-folds during the cosmologically relevant stage of
inflation can now be written as
\be \label{Nchi}
\frac{N[\chi]}{N} = -\int_{\chi}^{\chi_{end}}  \;
\frac{w(\chi)}{w'(\chi)} \; d\chi \;  \leqslant 1 \; .
\ee
In terms of the rescaled field and potential, the hierarchy of
slow roll parameters now becomes from eqs. (\ref{sig}) and (\ref{wchi}),
\bea
&&\epsilon_V  = \frac{1}{2N} \;
\left[\frac{w'(\chi)}{w(\chi)} \right]^2  \quad , \quad \eta_V
= \frac{1}{N}  \; \frac{w''(\chi)}{w(\chi)} \;  ,
\label{etav2} \\
&& \xi_V = \frac{1}{N^2} \; \frac{w'(\chi) \;
w'''(\chi)}{w^2(\chi)} \quad , \quad \sigma_V = \frac{1}{N^3}\;
\frac{[w'(\chi)]^2\,w^{(IV)}(\chi)}{w^3(\chi)} \; . \label{sig2}
\eea
It is clear from eqs. (\ref{uno}) and (\ref{etav2}) and (\ref{sig2})
that during the inflationary stage when wavelengths of
cosmological relevance cross the horizon $ w(\chi), w'(\chi) \sim
\mathcal{O}(1) $ and that this statement leads to a consistent slow
roll expansion as an expansion in $1/N$, namely the inverse of the
number of e-folds.

This equivalence between the slow roll and the $1/N$ expansion can
be made even more explicit by analyzing the Friedman equation and
the equation of motion for the inflaton in terms of the rescaled
field and potential. For this purpose, let us also introduce a
\emph{stretched}  (slow) dimensionless time variable $\tau$ and a
rescaled dimensionless Hubble parameter $h$ as follows
\be\label{time}
t = \sqrt{N} \; \frac{M_{Pl}}{M^2} \;  \tau \qquad
, \qquad H = \sqrt{N} \;  \frac{M^2}{M_{Pl}}\; h
\ee
in terms of
which the Friedman equation reads
\be \label{fredy}
h^2(\tau) = \frac13\left[\frac1{2\;N} \left(\frac{d\chi}{d \tau}\right)^2 +
w(\chi) \right]
\ee
and the evolution equation for the field
$\chi$ is given by
\be \label{eqnmot}
\frac1{N} \;  \frac{d^2
\chi}{d \tau^2} + 3 \; h \; \frac{d\chi}{d \tau} + w'(\chi) = 0
\ee
The slow-roll approximation follows by neglecting the
$\frac1{N}$ terms in eqs.(\ref{fredy}) and (\ref{eqnmot}). Both
$w(\chi)$ and $h(\tau)$ are of order $N^0$ for large $N$. Both
equations make manifest the slow roll expansion as an expansion in
$1/N$. The possibility of using $1/N$ as an expansion to study
inflationary dynamics was advocated previously in
ref.\cite{mangano}. The analysis above confirms this early
suggestion and establishes the slow roll expansion as a systematic
expansion in $1/N$.

In addition, at the absolute minimum of the potential $ w(\chi) ,
\; \chi_{min}$, one has to require $ w(\chi_{min}) =
w'(\chi_{min}) = 0 $ to guarantee that inflation ends after a
finite number of efolds. Moreover, we can choose $
|w''(\chi_{min})| = 1 $ without loss of generality. Then, the
inflaton mass around the minimum is given by a see-saw formula
\be
\label{mint} m = \frac{M^2}{M_{Pl}} \; .
\ee
This see-saw form of
the inflaton mass is again a hallmark of an effective field theory
and its smallness compared both to $M_{Pl}$ as well as $M\sim
10^{16}\textrm{GeV}$ is a consequence of the wide separation of
scales.

In particular the equation of motion
(\ref{eqnmot}) can be solved in an \emph{adiabatic} expansion in
terms of $1/N$, with the following result to zeroth order
\be \label{zerod}
\frac{d w(\chi) }{d\tau} = -
\frac{[w'(\chi)]^2}{3 \; h(\tau)} \left[ 1 + {\cal O}\left(\frac1{N} \right)
\right] \; ,
\ee
which again
requires for consistency that $w'(\chi) \sim w(\chi) \sim
\mathcal{O}(1)$ during slow roll. From eqs. (\ref{eqnmot})-(\ref{zerod})
it is clear that the slow field $\chi$ is a function
of the stretched (slow) time scale $\tau$.

We can now input the results from WMAP \cite{WMAP} to constrain the
scale $M$. The amplitude of adiabatic scalar perturbations in
slow-roll is expressed as
\be \label{ampliI}
|{\Delta}_{k\;ad}^{(S)}|^2  = \frac{1}{12 \, \pi^2 \;  M_{Pl}^6}
\; \frac{V^3}{V'^2}= \frac{N^2}{12 \, \pi^2} \;
\left(\frac{M}{M_{Pl}}\right)^4 \; \frac{w^3(\chi)}{w'^2(\chi)} \; ,
\ee
Since, $ w(\chi) $ and  $ w'(\chi) $ are of order one, we
find
\be\label{Mwmap}
\left(\frac{M}{M_{Pl}}\right)^2 \sim \frac{2
\, \sqrt{3} \, \pi}{N} \; |{\Delta}_{k\;ad}^{(S)}| \simeq  1.02
\times 10^{-5} \; .
\ee
where we used $ N \simeq 50 $ and the WMAP
value for $ |{\Delta}_{k\;ad}^{(S)}| = (4.67 \pm 0.27)\times
10^{-5} $ \cite{WMAP}. This fixes the scale of inflation to be
$$
M \simeq 3.19 \times 10^{-3} \; M_{PL} \simeq 0.77
\times 10^{16}\,\textrm{GeV} \; .
$$
This value pinpoints the scale of
the potential during inflation to be at the GUT scale suggesting a
deep connection between inflation and the physics at the GUT
scale in cosmological space-time.

That is, the WMAP data {\bf fix} the scale of inflation $ M $ for
single field potentials with the form given by eq.(\ref{wchi}).
This value for $M$ is below the WMAP upper bound on the inflation scale
$ 3.3 \; 10^{16}$GeV \cite{WMAP}.

Furthermore, the Hubble scale during (slow roll) inflation and the
inflaton mass near the minimum of the potential are thereby determined from
eqs.(\ref{time}) and (\ref{mint}) to be
\be\label{Hubsca}
 m = \frac{M^2}{M_{Pl}} = 2.45 \times 10^{13} \,\textrm{GeV}
\quad , \quad
H  = \sqrt{N} \; m \; h \sim 1.0 \times 10^{14}\,\textrm{GeV}
= 4.1 \; m \; .
\ee
since $ h = {\cal O}(1) $.

In absence of a underlying microscopic model from which   the
effective field theory description can be reliably extracted, we
can only \emph{surmise} the above form of the potential. Our main
observation is that the current phenomenological success of single
field slow-roll inflaton models, validated by the wealth of
observational data from WMAP \emph{strongly} suggests the
universal form (\ref{wchi}). Furthermore, such form yields a
  slow roll expansion consistently organized in powers of the
  small parameter $1/N$.

\subsection{A polynomial realization}

The results obtained above are very general and they only depend
on a recognition of the fast and slow fields and time scales
during the stage of slow roll inflation of cosmological relevance.
For specific models this general form severely constrains the
value of the couplings. In ref.\cite{nos} a thorough study of a
general quartic potential revealed that this simple
Ginzburg-Landau type effective field theory fits the WMAP data
remarkably well. Therefore, following ref.\cite{nos} we consider
the potential
\be\label{VI}
V(\phi)= V_0 \pm \frac{m^2}{2} \; \phi^2 +  \frac{ m
\; g }{3} \; \phi^3 + \frac{\lambda}{4}\; \phi^4 \; .
\ee
where $g, \; \lambda$ are dimensionless couplings.
The sign $+$ in the quadratic term corresponds to unbroken symmetry
while the $-$ sign describes the broken symmetry case.
We choose $ \lambda > 0 $ as a stability condition in order to
have a potential bounded from below while $ g $ may have any sign
(but we always have $ m^2 > 0 $).

The WMAP results rule out
the purely quartic potential ($m=0, \; g=0$). From the point of view
of an effective field theory it is rather \emph{unnatural} to set
$m=0$, since this is a particular point at which the correlation
length is infinite and the theory is critical. Indeed the
systematic study in ref.\cite{nos} shows that the best fit to the
WMAP data requires $m > 0$.

The general quartic Lagrangian eq.(\ref{V}) describes a
renormalizable theory. However, one can choose in the present context
arbitrary high order polynomials for  $ V(\phi) $.
These nonrenormalizable models
are also effective theories where $ M_{Pl} $ plays the r\^ole of UV cutoff.
However, already a quartic potential is rich enough to describe the full
physics and to reproduce accurately the data \cite{nos}.

Introducing the slow field $\chi$ as in eq. (\ref{wchi}) the
potential (\ref{VI}) can be written in the simple form
\be
V(\phi) = N \;  m^2 \;  M^2_{Pl} \left[w_0 \pm \frac{\chi^2}{2} +
\frac{G_3}{3} \;  \chi^3 + \frac{G_4}{4} \;  \chi^4 \right] \; ,
\label{VR}
\ee
where
\be w_0 = \frac{V_0}{N \; m^2 \;  M^2_{Pl}} \quad ,  \quad
G_3 = g \; \sqrt{N} \; \frac{M_{Pl}}{m}
\quad ,  \quad G_4 = \lambda \; N \; \left(\frac{M_{Pl}}{m}\right)^2
\ee
That is from eq.(\ref{wchi}),
$$
w(\chi)= w_0 \pm \frac12 \; \chi^2 + \frac{G_3}3 \; \chi^3 +
\frac{G_4}{4} \; \chi^4 \; .
$$
Comparing eqs.(\ref{wchi}) and (\ref{VR}) we read off the
relation (\ref{Hubsca})
between the inflaton mass $m$, the scale of the potential
$M$ and the Hubble parameter $H$  during slow roll inflation

Since slow roll inflation is consistently described with $w(\chi)
\sim \mathcal{O}(1) $, this in turn implies that $ G_3, \; G_4
\sim \mathcal{O}(1) $. This statement translates into the
following see-saw-like relations,
\be g = \frac{G_3}{\sqrt{N}}
\left( \frac{M}{M_{Pl}}\right)^2  \qquad ,  \qquad \lambda  =
\frac{G_4}{{N}} \left( \frac{M}{M_{Pl}}\right)^4
\label{lambdaG4} \; .
\ee
Therefore,  we {\bf naturally} find the order of
magnitude of the couplings to be:
\be  \label{natu}
g \sim 10^{-6} \quad
,  \quad \lambda \sim 10^{-12} \quad .
\ee
Since $ M/M_{Pl} \sim 3 \times 10^{-3} $,
these relations are a {\bf natural} consequence
of the validity of the effective field theory and of slow roll and
{\bf relieve} the fine tuning of the {\bf smallness} of the couplings.  
We emphasize that the
`see-saw-like' form of the couplings is a natural consequence of
the form of the potential (\ref{wchi}) and of the inequality
(\ref{Nchi}). While the hierarchy between the Hubble parameter,
the inflaton mass and the Planck scale during slow roll inflation
is well known, our analysis reveals that small couplings are
naturally explained in terms of powers of the ratio between the
inflationary and Planck scales \emph{and} integer powers of $ 1/\sqrt{N} $.

Therefore, one of our main results in this article, is that the
effective field theory and slow roll descriptions of inflation,
both validated by WMAP lead to conclude that there is no
fine tuning for the numerical values of the couplings. 
The smallness of the inflaton mass and the
coupling constants in this polynomial realization of the
inflationary potential is a \emph{direct} consequence of the
validity of both the effective field theory and the slow roll
approximations through a see-saw-like mechanism.

For a general potential $ V(\phi) $,
\be \label{serie}
V(\phi) = \sum_{n=0}^{\infty} \lambda_n \; \phi^n \; ,
\ee
we find from eq.(\ref{wchi})
\be\label{Gn}
\lambda_n =  \frac{G_n \; m^2}{\left( N \; M_{Pl}^2 \right)^{\frac{n}2-1}}
\quad {\rm where} \quad w(\chi) = \sum_{n=0}^{\infty} G_n \; \chi^n \; ,
\ee
and the dimensionless coefficients $ G_n $ are of order one.
We find the scaling behavior $\lambda_n \sim 1/N^{\frac{n}2-1}$.
Eq. (\ref{lambdaG4}) displays particular cases of eq.(\ref{Gn}) for $ n=3 $
and $ 4 $.

\bigskip

There are several remarkable features and consistency checks of the
relations (\ref{lambdaG4}):
\begin{itemize}
\item{Note the relation $\lambda \sim g^2$. This is the correct
consistency relation in a renormalizable theory because at one
loop level there is a renormalization of the quartic coupling (or
alternatively a contribution to the four points correlation
function) of orders $ \lambda^2 ,  \;  g^4 $ and $ \lambda \; g^2 $ which
are of the same order for $ \lambda \sim g^2 $. Similarly, at one
loop level there is a renormalization of the cubic coupling
(alternatively, a contribution to the three point function) of
orders $ g^3 $ and $ \lambda \; g $ which again require $ g^2 \sim
\lambda $ for consistency. }

\item {In terms of the effective field theory ratio $H/M_{Pl}$ and
slow roll parameters, the dimensionless couplings are\footnote{In
refs.\cite{pardec,qua} the cubic coupling $g$ corresponds to $m \;  g$
here with $ g $ dimensionless in the Lagrangian eq.(\ref{VI}). In
ref.\cite{pardec,qua} it was established that loop corrections
involve the ratio $ g/H $ in the notation of that reference.}
\be
\frac{m \;  g}{H}  \sim  \frac{1}{N} \;
\frac{H}{M_{Pl}}  \qquad , \qquad
\lambda  \sim  \frac{1}{N^2} \, \left(\frac{H}{M_{Pl}} \right)^2
\label{lam} \; .
\ee
These relations agree with those found for the
dimensionless couplings in ref.\cite{pardec,qua} once the slow
roll parameters are identified with the expressions
(\ref{etav2})-(\ref{sig2}) in terms of $1/N$.  The results of
refs.\cite{pardec,qua} revealed that the loop expansion is indeed an expansion
in the effective field theory ratio $ H/M_{Pl} $ and the slow roll
parameters. Our study here allows us to go further and state that
the loop expansion is a consistent double series in the effective
field theory ratio $ H/M_{Pl}$ \emph{and} $1/N$. Loops are either
powers of $ g^2 $ or of $ \lambda $ which
implies that for each loop there is a factor $ H^2/M^2_{Pl} $. The
counting of powers of $1/N$ is more subtle: the nearly scale invariant
spectrum of fluctuations leads to  infrared enhancements of quantum
corrections in which the small factor $1/N$ enters as an infrared
regulator. Therefore large denominators that feature the infrared
regulator of order $1/N$ cancel out factors $1/N$ in the
numerator. The final power of $1/N$  must be computed in
detail in each loop contribution. }

\item{We find the relation (\ref{lambdaG4})to be very suggestive.
Since the scale of inflation $M$ is fixed, presumably by the
underlying microscopic (GUT) theory, the scaling of $\lambda$
with the inverse of the number of e-folds strongly suggests a
\emph{renormalization group explanation of the effective field
theory} because the number of e-folds is associated with the
logarithm of the scale $ N=\ln a $. A renormalization group
improved scale dependent quartic coupling behaves \cite{weipes} as
$ \lambda(K) \propto 1/\ln K $ with $ K $ the scale at which the
theory is studied. Since in an expanding cosmology the physical
scale grows with the scale factor it is natural to expect that a
renormalization group resummation program would yield that the
renormalized coupling scales as
$$
\lambda \sim 1/\ln a  \sim 1/N \; .
$$
This of course requires further study.  }
\end{itemize}

\subsection{Gauge Invariant Scalar Perturbations}

Let us now see that the slow roll approximation appears as a
consistent expansion in $ 1/N $ in the mode equations
for the gauge invariant scalar perturbations  $ u_k({\eta}) $.
These can be written as \cite{hu} \be\label{ecmu} \left[
\frac{d^2}{d {\eta}^2} + k^2 -\frac1{z} \; \frac{d^2 z}{d
{\eta}^2}\right]u_k({\eta})=0 \; , \ee where $ {\eta} $ is the
conformal time related, as usual, to  cosmic time $  t $ by $ d {
t} = a({ t}) \; d{\eta} $ and $ z({\eta}) $ stands for
$$
z({\eta}) \equiv a^2 \; \frac{d \phi}{d a} \; .
$$
In terms of the the slow and dimensionless variables
$$
\bar{\eta} \equiv \eta \; \sqrt{N} \; m  \quad , \quad
\chi= \frac{\phi}{M_{Pl} \; \sqrt{N}}\quad , \quad
h = \frac{H}{m \; \sqrt{N}} \quad , \quad
\bar{k}= \frac{k}{m \; \sqrt{N}} \quad  ,
$$
eq.(\ref{ecmu}) takes the form
\be\label{ecmu2}
\left[ \frac{d^2}{d {\bar \eta}^2} + {\bar k}^2 -\frac1{z} \;
\frac{d^2 z}{d {\bar\eta}^2}\right]u_k({\bar\eta})=0 \quad  .
\ee
We can compute the `potential' in eq.(\ref{ecmu2}) in terms of $
w(\chi) $ and its derivatives using the evolution equations
(\ref{eqnmot}). We find to first order in $1/N$,
$$
\frac1{z} \;  \frac{d^2 z}{d {\bar \eta}^2} = \frac2{{\bar\eta}^2}
\left\{ 1 + \frac{3}{2N} \left[ -\frac{
w''(\chi)}{ w(\chi) } + \frac{3 \;  w'^2(\chi)}{2 \; w^2(\chi) }
\right] + {\cal O}\left( \frac1{N^2} \right) \right\}
$$
where we used the following expression for the dimensionless
conformal time
$$
\bar\eta = - \frac1{h \, a} \left[ 1 +\frac{1}{ 2 \; N} \;
\frac{w'^2(\chi)}{w^2(\chi) } + {\cal O}\left( \frac1{N^2}
\right) \right]
$$
Eq.(\ref{ecmu2}) takes the  familiar form \cite{hu} in terms of the above
dimensionless variables,
\be \label{ecmod}
\left\{ \frac{d^2}{d {\bar\eta}^2} + {\bar k}^2
-\frac2{{\bar\eta}^2}\left[1 + \frac12(9 \; \epsilon_V - 3 \; \eta_V)  +
{\cal O}\left( \frac1{N^2} \right)\right] \right\}u_k({\bar\eta})=0
\; ,
\ee
and explicitly exhibiting [see eq.(\ref{etav2})], once again, that the
slow roll approximation is an expansion in $ 1/N $.

Relevant modes for the large scale structure and the CMB are today
in the range from $ 0.1$ Mpc to $103$ Mpc. These scales at the
beginning of inflation correspond to physical wavenumbers in the
range
$$
e^{N_T-60} \; 10^{16} \, GeV < k < e^{N_T-60} \; 10^{20} \, GeV
$$
where $N_T$ stands for the total number of efolds (see for example
Ref. \cite{sd}). Therefore, $ \bar k \gg 1 $ and eq.(\ref{ecmod}) is deep
in the semiclassical regime.

\section{The tensor/scalar ratio $r$ and  Lyth's bound clarified}

The next step in CMB observations is the search for B-modes which
if observed can place a direct bound on the scale of inflation.
The measurement of B-modes or tensor perturbations with CMB
experiments depends on the magnitude of the ratio $r$ between
tensor and scalar perturbations.

As noticed by Lyth \cite{lyth} the ratio of tensor/scalar fluctuations
$ r $ can be related to the change of the inflaton field while
the cosmologically relevant modes exit the horizon by
\be \label{ly}
\frac{\Delta \; \phi}{M_{Pl}} \sim \sqrt{\frac{r}8} \; \;\Delta N \; .
\ee
For $ \Delta N \simeq 4 $ this gives
\be \label{ly2}
\frac{\Delta \phi}{M_{Pl}} \sim \sqrt{2 \; r} \; .
\ee
A more stringent bound has been found in ref.\cite{efs} by a statistical
analysis of over $2\times 10^6$ slow roll inflationary models
\be \label{efsbound}
\frac{\Delta \phi}{M_{Pl}} \sim 6 \; \sqrt{8 \; \pi} \;
r^{\frac{1}{4}}  \; .
\ee
Inflationary model building proposes an effective field theory
description of the inflationary potential of the
form \cite{lyth,vpj}
\be\label{Vb}
V(\phi) =  V_0 + \frac{m^2}{2} \,\phi^2 + \phi^4 \;
\sum_{p=0}^{\infty}\lambda_p  \; \left(\frac{\phi}{M_P}\right)^p  \; .
\ee
Within this framework, it is often stated that the validity of the
effective field theory description entails that
\be
\frac{\phi}{M_{Pl}} \ll 1 \label{val}  \; .
\ee
A tension between this stringent constraint on the validity of an
effective field theory and the bounds (\ref{ly2}) or
(\ref{efsbound}) is evident. The validity of the constraint
(\ref{val}) entails that $ \Delta \phi / M_{Pl} \ll 1 $ suggesting
that effective field theory predicts values of $ r \ll 1 $  probably
too small to be observed in the next generation of CMB
experiments \cite{vpj}. Alternatively if larger values of $r$ are
measured then this would entail a breakdown of the effective field
theory approach to inflation.

We find this line of reasoning incorrect for three different but
complementary reasons:

\begin{itemize}
\item{The validity of an  effective field theory expansion does
\emph{not} rely on  $\phi/M_{Pl} \ll 1$, instead it relies on a
wide separation between the scale of inflation and the higher
energy scale of the underlying microscopic theory. If the
effective field theory emerges from integrating out degrees of
freedom at the GUT scale, then $ H/M_{GUT} \sim 10^{-2} $, if such
scale is instead the Planck scale then $ H/M_{Pl} \sim 4 \times
10^{-5} $, and in either case an effective field theory
description is valid. Indeed, detailed calculations in
ref.\cite{pardec,qua} reveal that the quantum corrections to
slow-roll inflation are of the order $ \left(H/M_{Pl}\right)^2 $.
Any breakdown of an effective field theory is typically manifest
in large quantum corrections but the results of \cite{pardec,qua}
unambiguously point out that quantum corrections are well under
control for $H/M_{Pl} \ll 1$. This provides a reassuring
confirmation of the validity of the effective field theory for a
scale of inflation consistent with the WMAP data. }

\item{Three simple examples highlight that the criterion
$\phi/M_{Pl} \ll 1$ \emph{cannot} be the deciding factor for the
reliability of the effective field theory: consider the following
series
\bea & & \sum_{p=0}^{\infty}(-1)^p \left(\frac{\phi}{M_{Pl}}
\right)^{2 \; p} = \frac{1}{1+ \left(\frac{\phi}{M_{Pl}}\right)^2}
\label{seriesA} \\
& & \sum_{p=0}^{\infty}\frac{(-1 )^p}{(2p)!}\,\left(\frac{\phi}{M_{Pl}}
\right)^{2 \; p} = \cos\left(\frac{  \phi}{M_{Pl}}
\right)\label{seriesB}\\
& & \sum_{p=0}^{\infty} \left(\frac{\phi}{M_{Pl}} \right)^{2 \; p}
=  \frac{1}{1- \left(\frac{\phi}{M_{Pl}}\right)^2} \label{seriesC}
\; . \eea The sum of both series (\ref{seriesA}) and
(\ref{seriesB}) is perfectly well defined for $\phi > M_{Pl}$. In
particular, the series (\ref{seriesB}) is a prototype for an
axion-type potential \cite{natu}, while certainly the series
(\ref{seriesC}) breaks down for $\phi \sim M_{Pl}$. The series
(\ref{seriesA})-(\ref{seriesB}) do not feature any real
singularity in the variable $\phi/M_{Pl}$ whereas
eq.(\ref{seriesC}) has a singularity at $\phi = M_{Pl}$. These
elementary examples highlight that what constraints the
reliability of the effective field theory description of the
inflationary potential is \emph{not} the value of the ratio
$\phi/M_{Pl}$ but the position of the \emph{singularities} as a
function of this variable. These singularities are determined by
the large order behavior of the \emph{coefficients} in the series.
If the series has a non-zero radius of convergence or if it is
just Borel summable, it defines the function $ V(\phi) $ {\bf
uniquely}. Clearly, a thorough knowledge of the radius of
convergence of the series in the effective field theory requires a
detailed knowledge of the underlying microscopic theory. However,
it should also be clear that the requirement $\phi/M_{Pl}\ll 1$ is
overly restrictive in general. } \item{One of the main results of
this article is that the combination of WMAP data and slow roll
expansion suggest a universal form of the inflation potential, \be
V(\phi) = N \; M^4 \;  w(\chi) \quad , \quad \chi =
\frac{\phi}{\sqrt{N} \; M_{Pl}} \quad . \ee Even in the case when
the coefficients in a $\chi$-series expansion of $ w(\chi) $ lead
to a breakdown of the series at $ \chi \sim 1 $, namely at $\phi
\sim \sqrt{N} \; M_{Pl}$, there is still room for values of $
M_{Pl} < \phi < \sqrt{N} \;  M_{Pl} $ for which the series would
be reliable. However, no \emph{ a priori} physical reason for such
a breakdown can be inferred without a reliable calculation of the
effective field theory from a microscopic fundamental theory.
Therefore, we expect that the effective field theory potential
$V(\phi)=N \; M^4 \; w(\chi)$ would be reliable \emph{at least} up
to $\phi \sim \sqrt{N} \; M_{Pl}$ and most generally for values $
\chi \gg 1 $ and hence $ \phi \gg M_{Pl} $. }
\end{itemize}

As mentioned above, the studies in ref.\cite{pardec,qua} reveal
that  quantum corrections in the effective field theory yields an
expansion in $ \left(\frac{H}{ M_{Pl}}\right)^2 $ for
\emph{general inflaton potentials}. This indicates
that the use of the inflaton potential $ V(\phi) $ from effective
field theory is consistent for
$$
 \left(\frac{H}{M_{Pl}}\right)^2 \ll 1 \quad {\rm and~hence} \quad
V(\phi) \ll  M_{Pl}^4 \; .
$$
We  find using eq.(\ref{wchi}): \be w(\chi) \ll \frac1{N} \;
\left(\frac{M_{Pl}}{M}\right)^4 \sim \frac1{N} \; 10^{12} \; . \ee
This condition yields an  {\bf upper bound} in the inflaton field
$ \phi $ depending on the large field behaviour of $ w(\chi) $. We
find for relevant behaviours of the inflaton potential the
following upper bounds on $ \chi $ and $ \phi$:
\bea \label{cotxi}
&&  w(\chi) \buildrel{\chi \gg 1}\over \simeq \chi^2 \quad :
 \quad \chi \ll \frac{10^6}{\sqrt{N}} \quad , \quad
\phi \ll  10^6 \; M_{Pl} \cr \cr
&& w(\chi) \buildrel{\chi \gg 1}\over \simeq \chi^4 \quad :
 \quad \chi \ll \frac{10^3}{N^{\frac14} }
 \quad , \quad  \phi \ll 2659 \; M_{Pl} \quad {\rm for} \quad N\simeq 50
\cr \cr
&& w(\chi) \buildrel{\chi \gg 1}\over \simeq \chi^n \quad :
 \quad \chi \ll \left(\frac{10^{12}}{N}\right)^{\frac1{n}}
 \quad , \quad  \phi \ll 10^{\frac{12}{n}} \; N^{\frac12 - \frac1{n}}
\; M_{Pl} \cr \cr
&& w(\chi) \buildrel{\chi \gg 1}\over \simeq e^{\chi}\quad :
 \quad \chi \ll 12 \; \ln 10 - \ln N = 23.72
 \quad , \quad  \phi \ll 167 \;  M_{Pl}\quad {\rm for} \quad N\simeq 50 \; .
\eea We see that the effective field theory is {\bf consistent}
for values of the inflaton field well {\bf beyond} the Planck mass
even for very steep potentials, such as the exponential function $
e^{\chi} $.

\bigskip

We summarize this discussion  with  the following two remarks,
which when taken together, relieve the tension between the
accessible experimental values of $r$ and the validity of the
effective field theory description:

\begin{itemize}
\item{The inflaton potential $ V(\phi) $ may be applicable even
for large values of  $ \frac{\phi}{M_{Pl}} $ {\bf invalidating }
the arguments leading to the bounds
eqs.(\ref{ly2})-(\ref{efsbound}) for $ r $. The  WMAP data
constrains  $ r \lesssim 0.16 $ \cite{WMAP}, the true value for $ r
$ may be very well much smaller than the present upper bound.}

\item{If for some reason (or prejudice) one wishes to restrict the
analysis to values of the field where  $ \chi \lesssim 1 $, then
we provide from our study the following {\bf improved} bounds:
\bea\label{cot1}
&& \Delta\chi = \frac{\Delta \phi}{\sqrt{N} \; M_{Pl}} \sim
\sqrt{\frac{2 \; r}{N}} \quad {\rm for ~ the ~ bound ~in ~ref.}
 [17] \cr \cr && \Delta\chi =\frac{\Delta \phi}{\sqrt{N} \;
M_{Pl}} \simeq \frac6{\sqrt N} \; \sqrt{8 \, \pi} \; r^{\frac14}
\quad
{\rm for~ the~ bound~ in ~ref.}[18] \; .
\eea
This gives for $ \Delta\chi \sim 1 $, the {\bf improved} bounds on $ r $:
\bea
&& r \lesssim \frac{N}2 \simeq 25 \quad {\rm for~ the~ bound~ in
 ~ref.} [17] \cr \cr
&&  r \lesssim \frac{N^2}{6^4 \; (8\,\pi)^2} \simeq 0.003 \quad
{\rm for~ the~ bound~ in ~ref.}[18] \quad   {\rm and} \quad
N\simeq 50 \; . \eea It must be noticed that these bounds depend
on whether one uses as Planck's mass $ M_{Pl} = 1/\sqrt{8  \pi G}
$ or $ m_{Pl} = 1/\sqrt{G} $, $G$ being Newton's constant. Here we
have used the first definition $M_{Pl}$ as in ref.\cite{lyth}.
Ref. \cite{efs} uses the second definition $ m_{Pl} $.

We find using $ m_{Pl} = 1/\sqrt{G} $
\bea
&& \frac{\Delta\chi}{\sqrt{8\, \pi}} = \frac{\Delta \phi}{\sqrt{N} \; m_{Pl}}
\sim \frac12 \; \sqrt{\frac{r}{\pi \; N}}
\quad {\rm for~ the~ bound~ in ~ref.} [17] \cr \cr
&& \frac{\Delta\chi}{\sqrt{8\, \pi}} =\frac{\Delta \phi}{\sqrt{N} \; m_{Pl}}
\simeq \frac6{\sqrt N} \; r^{\frac14} \quad
 {\rm for~ the~ bound~ in ~ref.}[18] \; .
\eea
And we then find for
$ \frac{\Delta\chi}{\sqrt{8\, \pi}} \sim 1 $:
\bea\label{cot4}
&& r \lesssim 4 \; \pi \; N \simeq 628 \quad {\rm for~ the~ bound~ in
 ~ref.} [17] \cr \cr
&&  r \lesssim \frac{N^2}{6^4} \simeq 1.93 \quad
{\rm for~ the~ bound~ in ~ref.}[18] \quad   {\rm and} \quad N\simeq 50
\; .
\eea}
\end{itemize}

In conclusion, even under the more conservative assumption
$\chi \lesssim 1$, we provide the {\bf improved}  bounds
eqs.(\ref{cot1})-(\ref{cot4}) which allow  values of $ r $ {\bf
substantially larger} than the original ones \cite{lyth,efs}. This
is  a consequence of the  factor $ \sqrt{N} \sim 7 $ in our slow
variable $ \chi $, which in turn is the result of a consistency
between the WMAP data and slow roll in an effective field theory
description. The structure eq.(\ref{wchi}) of the inflaton
potential therefore relieves the tension between the values of  $
\frac{\phi}{M_{Pl}} $ and $ r $.

We remark that the arguments presented above suggest that the
reluctance to use the inflaton potential $ V(\phi) $ for $ \phi
\gtrsim M_{Pl} $ arises from a prejudice which is unwarranted
under the most general circumstances, unless of course, the
inflaton potential features singularities. The true upper bound
for the validity  of the effective field theory description of
inflation is $ V(\phi) \ll M_{Pl}^4$ which in fairly general cases
allows large values of $ \frac{\phi}{M_{Pl}} $ as emphasized by
eqs. (\ref{cotxi}).

\section{Conclusions}

In this article we show that the consistency of an effective field
theory description of inflation with the WMAP data and the
slow roll approximation provide a universal form of the
inflationary potential. This form leads to a clear understanding
of the validity of the effective field theory and makes manifest
that the slow roll expansion is an expansion in $1/N$ where $N\sim
50$ is the number of e-folds before the end of inflation when
cosmologically relevant scales exit the Hubble
radius. The inflaton potential is of the form,
$$
V(\phi) = N \; M^4 \;  w(\chi)\; ,
$$
\noindent where the WMAP data pinpoints $M$ at the GUT scale $M\sim
0.77\times 10^{16}~\textrm{GeV}, \; \chi$ is a slowly varying
dimensionless field,
$$
\chi = \frac{\phi}{M_{Pl} \; \sqrt{N}} \; ,
$$
\noindent and $w(\chi) \sim \mathcal{O}(1)$.

The dynamics of the field $\chi$ depends solely on the
\emph{stretched}  (slow) time variable
$$
\tau = \frac{t \;  M^2}{M_{Pl} \; \sqrt{N}}  = \frac{m \;t}{\sqrt{N}} \; ,
$$
\noindent and is determined by the equation of motion (\ref{eqnmot}) which
can be solved consistently in an expansion in $1/N$.

This form of the potential makes explicit  that the slow roll
expansion is a consistent expansion in $1/N$, see eqs.
(\ref{etav2})-(\ref{sig2}). This also shows up in
the equations of motion for the mode functions of
gauge invariant scalar perturbations.

\medskip

A polynomial realization of the inflaton potential as an effective
field theory of the Ginzburg-Landau form, which has recently been
shown to fit the WMAP data remarkably well \cite{nos}, indicates that
the Hubble parameter, the inflaton mass and the  non-linear
couplings emerge as powers of the see-saw-like ratio $ M/M_{Pl}\sim
3 \times 10^{-3} $. The \emph{smallness} of which warrants the
validity of the effective field theory. Thus, it is clear that the
smallness of the non-linear self-couplings is {\bf not} a result of
fine tuning, but a  {\bf natural} consequence of an effective field
theory in which self-couplings emerge from see-saw like mechanisms
with two widely different scales: the inflation (or GUT) and Planck
scales. Furthermore, the consistency between the slow roll and $1/N$
expansions implies that the quartic self-coupling $ \lambda $ scales
as $ \lambda \sim 1/N \sim 1/\ln a $ and that the cubic
self-coupling $ g $ and the quartic self-coupling are such that $
g^2 \sim \lambda $.

\medskip

Our observations and results here relieve the tension between the
validity of the effective field theory approach and the values of
the ratio $r$ between tensor and scalar perturbations. This
tension is actually an \emph{artificial} result of a
\emph{prejudice} on the validity of the effective field theory.
This validity is not determined by the  maximum  value of
$\phi/M_{Pl}$ but rather on the smallness of the ratio $H/M_{Pl}$.
These arguments pave the way for an \emph{unprejudiced}
observational exploration of B-modes in the next generation of CMB
experiments \emph{within} the theoretical description of slow roll
inflation based on an effective field theory.

\medskip

The effective field theory describing slow roll inflation during
the stage relevant for cosmology features remarkable properties
which indicate that inflation is
hovering \emph{near} a trivial gaussian infrared fixed point in the
renormalization group sense. Three important aspects are behind
this conjecture:

i) the nearly scale invariant power spectrum of
scalar perturbations,

ii) the fact that the coupling constant
associated with a dimension four operator, $ \lambda $ (the quartic
coupling)  scales with the scale factor  as $ \lambda \sim
1/\ln a $ and

iii) the fact that the quantum corrections \cite{pardec,qua}
are in terms of the effective field theory ratio $ H/M_{Pl} $ and
powers of $ 1/\ln a $.

This behavior is similar to that of a renormalizable field theory
near its trivial fixed point.   We will continue to explore this
remarkable aspect of slow roll inflation and expect to report on
these studies in the future.

\bigskip

We can summarize our main present results as follows: we trade the
small parameters in the inflationary models for appropriate slow
variables by introducing two crucial and independent ingredients:

\begin{itemize}
\item{ first, by introducing the  inflationary or GUT energy scale
$M \sim 10^{16}$GeV and the Planck scale  $M_{Pl}$ in the inflaton
field and in the inflaton potential. }

\item{second, by rescaling the inflaton field with the square root of the
number of efolds $\sqrt{N}$. This turn to introduce a dependence of
the couplings on $ N $ similar to a renormalization group running of
the couplings. }
\end{itemize}

\begin{acknowledgments}   We thank Hiranya Peiris and Olivier Dor\'e
for useful discussions about section III during the 9th Paris
Chalonge Cosmology Colloquium. D.B. thanks the US NSF for support
under grant PHY-0242134,  and the Observatoire de Paris and LERMA
for hospitality during this work. This work is supported in part
by the Conseil Scientifique de l'Observatoire de Paris through an
`Action Initiative, BQR'.
\end{acknowledgments}

\end{document}